\documentclass[12pt]{iopart}
\usepackage{iopams}
\usepackage{graphicx}
\usepackage{setstack}
\usepackage{bm}
\usepackage{srctex}
\usepackage{hyperref}

\begin{document}
\title[Continuous transformation between FM and AFM circular structures]{Continuous transformation between ferro and antiferro circular structures in $J_1-J_2-J_3$ frustrated Heisenberg model}

\author{V E Valiulin$^{1,2,3}$, A V Mikheyenkov$^{1,2,3}$,
N M Chtchelkatchev$^{1,2,4}$ and A F Barabanov$^2$}

\address{$^1$ Institute for High Pressure Physics, Russian Academy of Sciences, Moscow (Troitsk) 108840, Russia}
\address{$^2$ Department of Theoretical Physics, Moscow Institute of Physics and Technology  (State University), Moscow 141700, Russia}
\address{$^3$ National Research Centre ``Kurchatov Institute'', Moscow 123182, Russia}
\address{$^4$ Ural Federal University, Ekaterinburg 620002, Russia}
\ead{n.chtchelkatchev@gmail.com}

\begin{abstract}
Frustrated magnetic compounds, in particular low-dimensional, are topical research due to persistent uncover of novel nontrivial quantum states and potential applications. The problem of this field is that many important results are scattered over the localized parameter ranges, while areas in between still contain hidden interesting effects. We consider $J_1-J_2-J_3$ Heisenberg model on the square lattice and use the spherically symmetric self-consistent approach for spin-spin Green's functions in ``quasielastic'' approximation. We have found a new local order in spin liquids: antiferromagnetic isotropical helices. On the structure factor we see circular concentric dispersionless structures, while on any radial direction the excitation spectrum has ``roton'' minima. That implies nontrivial magnetic excitations and consequences in magnetic susceptibility and thermodynamics. On the $J_1-J_2-J_3$ exchange parameters globe we discover a crossover between antiferromagnetic-like local order and ferromagnetic-like; we find stripe-like order in the middle. In fact, our ``quasielastic'' approach allows investigation of the whole $J_1-J_2-J_3$ globe.
\end{abstract}
\submitto{\JPCM}
\maketitle

\section{Introduction}\label{sec.intro}

One of the key topical questions in the field of magnetism is how strong frustration coexists with ordering.~\cite{Balent10_N,Kawamu11_JPCS,Seabra16_PRB,Danu16_PRB,Hu17_SR,Ferrar17_PRB,Buesse18_PRL,Bishop18_JPCS}
Intense research today addresses systems with multiple
frustrating mechanisms. The problems is how the number of frustrating
mechanisms and the relations between them affect the order and
the structure of the disordered state.
The theoretical activity in the field is continuously fed by
regular experimental achievements.
New possibilities to construct and control quantum states of matter emerge this way including transport of skyrmions and antiskyrmions,~\cite{McGrou16_NJP,Zhang17_NC,Sutcli17_PRL,Leonov17_NC,Liang18_NJP}
chiral spin liquids with robust edge modes,~\cite{Hickey17_PRB,Claass17_NC}
nontrivial quasiparticles like semions.~\cite{Yao18_NP}

Frustration mechanisms in magnetic systems have different nature, including magnetoelastic coupling,~\cite{Wahish17_PRB} spin-orbital interaction,~\cite{Belemu14_PRA,Belemu17_PRB,Mikhee18_JETP,Belemu18_NJP} geometrical constraints,~\cite{Wietek17_PRB,Ling17_PRB,Manson18_SR} doping, competing interactions (both exchange~\cite{Chandr90_PRL,Chubuk91_PRB,Siurak01_PRB,Mambri06_PRB,Jiang12_PRB,Bauer17_PRB, Buesse18_PRL, Ferrar17_PRB, Iqbal18_a} and long-range order --- Dzyaloshinskii-Moriya~\cite{Huang16_NJP,Messio17_PRL} and dipole-dipole~\cite{Ye17_PRB} ones).

There is a wide class of magnetically frustrated systems that can be well enough described as a set of weakly interacting magnetic square lattice planes with strong multiexchange Heisengerg interaction within the plane.
This concept is during decades widely used for the spin system of HTSC cuprates~\cite{Tranqu07_BookChap,Plakid10_BookChap}
and for long known other layered compounds.~\cite{Melzi00_PRL,Melzi01_PRB,Rosner03_PRB,Kageya05_JPSJ,Vasala14_JPCM}
Later several other layered (quasi-two-dimensional) $J_1-J_2$ compounds
were discovered covering a great variety of relationships between first and second exchange parameters. In particular, these are  Pb$_2$VO(PO$_4$)$_2$,~\cite{Kaul04_JMMM,Skoula07_JMMM,Carret09_PRB,Skoula09_EL}
(CuCl)LaNb$_2$O$_7$~\cite{Kageya05_JPSJ}, SrZnVO(PO$_4$)$_2$,~\cite{Skoula09_EL,Tsirli09_PRBa,Tsirli09_PRB,Bosson10_PRB}
BaCdVO(PO$_4$)$_2$,~\cite{Carret09_PRB,Tsirli09_PRBa,Nath08_PRB}
K$_2$CuF$_4$, Cs$_2$CuF$_4$, Cs$_2$AgF$_4$, La$_2$BaCuO$_5$, Rb$_2$CrCl$_4$,~\cite{Feldke95_PRB,Feldke98_PRB,Manaka03_PRB,Kasina06_PRB,Tsirli09_PRB,Carret09_PRB,Tsirli13_PRB}
and others.

Today multi-exchange, in particular $J_1$-$J_2$-$J_3$ strongly frustrated low-dimensional Heisenberg systems
are in the centre of attraction due to the progress of material science, development of new theoretical tools and new physics emerging from
competition of $J$-frustrating mechanisms.~\cite{Sindzi09_JPCS,Sindzi10_JPCS,Feldne11_PRB,Reuthe11_PRB, Danu16_PRB,Iqbal16_PRB,Hu17_SR,Bauer17_PRB,Ferrar17_PRB,Oitmaa17_PRB,Messio17_PRL,
Sapkot17_PRL,Schect17_PRL,Tymosh17_PRX,Wang17_a,Wietek17_PRB,Ye17_PRB,Buesse18_PRL,Iqbal18_a}

The problem of this field is that many important results are scattered over the localized parameter ranges, while areas in between still contain hidden new effects. We use the approach~\cite{Kondo72_PTP,Shimah91_JPSJ,Baraba94_JPSJ,Hartel13_PRB,Mikhey16_JMMM} which provides an opportunity to uncover ``white spots'' on $J_1-J_2-J_3$-``globe''.

Conventionally, dealing with the phase diagram of the $J_1-J_2-J_3$ Heisenberg model at zero external magnetic field, we can fix the length of the vector $\mathbf J=(J_1,J_2,J_3)$ thus referring to the globe picture.

We investigate how local order continuously evolves in spin liquids between antiferromagnetic~\cite{Reuthe11_PRB, Iqbal16_PRB} and ferromagnetic~\cite{Seabra16_PRB} isotropical helices. On the structure factor we observe evolution of the circular concentric dispersionless structures originating from quantum fluctuations, while on any radial direction the excitation spectrum has ``roton'' minima. That implies nontrivial magnetic excitations and consequences in magnetic susceptibility and thermodynamics. On the $J_1-J_2-J_3$ exchanges globe we pay special attention to the crossover between antiferromagnetic-like (AFM) local order and ferromagnetic-like (FM); we find a stripe-like order in the middle of this crossover.

By now only the domain $J_2>0$, $J_3=0$ (that is, half of the globe equator) can be considered as deeply investigated, see, e.g., Refs.~\cite{Balent10_N,Li12_PRB,Hartel13_PRB,Mikhey16_JMMM,Wang18_PRL,Mikhee18_JETP} and references therein. Briefly, the generally accepted picture is the following. At $T=0$ for $J_1>0$ there are two phase transitions in the system: from AFM long-range order to spin liquid and then to stripe-like long-range order. For $J_1<0$ there is a sequence of transitions: stripe -- spin liquid -- FM order~\cite{Shanno04_EPJB,Shanno06_PRL,Sindzi07_JMMM,
Schmid07_JPCM,Schmid07_JMMM,Viana07_PRB,Sindzi09_JPCS,
Shindo09_PRB,Hartel10_PRB,Hartel13_PRB,Dmitri97_PRB,Richte14_JPCS,Ren14_JPCM}. At the nonzero temperature the same applies to the short-range order structure.

Still, there is no full clarity on the nature of successive quantum phase transitions, fine details of the disordered state, influence of finite temperature (at least in quasi-two-dimensional case) and nonzero $J_3$.

The ``quasielastic'' approach adopted here allows to resolve or dampen the mentioned problems. In particular, it is possible to investigate the whole $J_1-J_2-J_3$ globe. We can find out spin-spin Green's and correlation functions, structure factor, correlation length (also spin susceptibility and heat capacity) in the wide temperature and exchange parameters range. Our semianalytical calculation method is accurate enough; so, we reproduce quantitatively the results obtained numerically in Refs.~\cite{Seabra16_PRB,Reuthe11_PRB, Iqbal16_PRB} as discussed in detail below in Concussions.


\begin{figure}[t]
  \centering
  \includegraphics[width=.4\columnwidth]{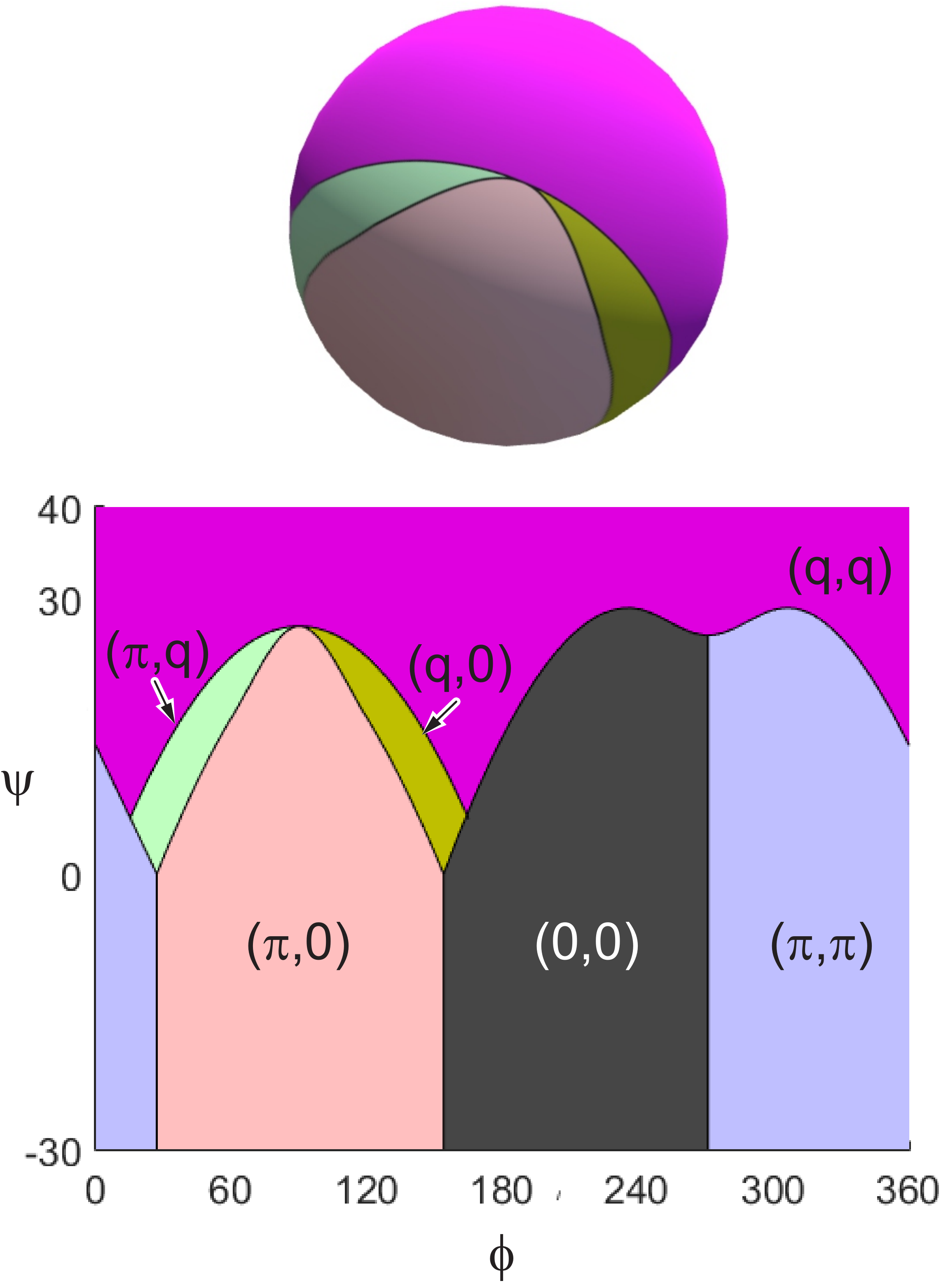}\\
  \caption{(Color online) A sketch of the $J_1-J_2-J_3$-model phase diagram in the classical limit.  The labels mark the positions of structure factor $\delta$-peak, see text and Eqs.~\ref{S_r}, \ref{a_Cq_1}. \textit{Top}: ``Globe'' representation of the phase diagram when $J_1 = \cos(\psi)\cos(\phi)$, $J_2 = \cos(\psi)\sin(\phi)$, $J_3 = \sin(\psi)$, see Sec.~\ref{Phase_diagram} (hereafter all the energy quantities are normalized by $\sqrt{(J_1^2+J_2^2+J_3^2)}$).
  \textit{Bottom}: ``Flat'' representation of the phase diagram.
  The phases are: $(0,0)$ --- ferromagnetic (FM), $(\pi,\pi)$ --- antiferromagnetic (AFM), $(\pi,0)$ --- stripe, while $(\pi,q)$, $(q,0)$ and $(q,q)$ are three different incommensurate helical phases.
  We concentrate here on the helical structures, so accordingly we have chosen the visible side of the globe. Thus, the FM and AFM states are on the dark side of the globe and not visible.
  To avoid confusion we measure parametrical angles $\varphi$ and $\psi$ in degrees and the Brillouin zone coordinates – conventionally in radians.
  }\label{Globe}
\end{figure}
\section{Multi-exchange Heisenberg system: from simple frustration to quantum helices}\label{sec.multi}

\begin{figure}[b]
  \centering
  \includegraphics[width=.25\columnwidth]{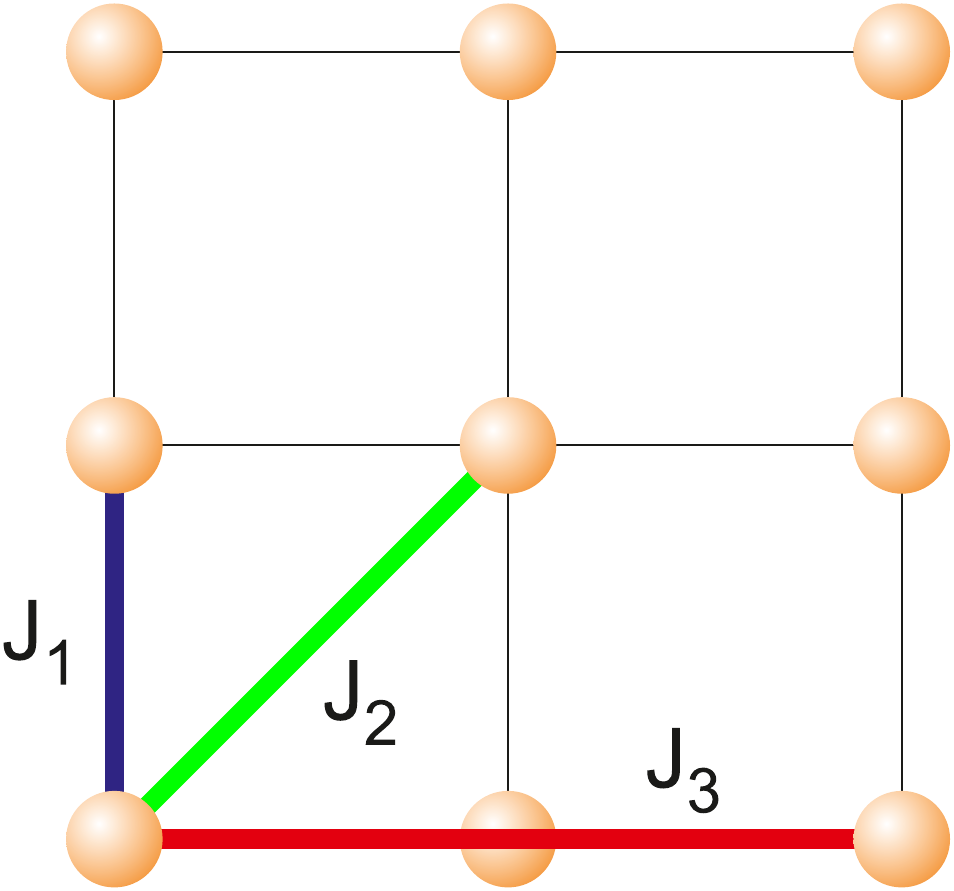}\\
  \caption{(Color online) The sketch of the square lattice and three exchange bonds.
  }\label{Square}
\end{figure}
\subsection{Model Hamiltonian}\label{sec.hamon}

We address two-dimensional $J_1-J_2-J_3$ Heisenberg model with spin $S=1/2$ on the square lattice, see Fig.~\ref{Square}. The Hamiltonian of the model reads
\begin{equation}
H=J_{1}\sum_{\langle\mathbf{i},\mathbf{j}\rangle}
\widehat{\mathbf{S}}_{\mathbf{i}}\widehat{\mathbf{S}}_{\mathbf{j}}+
J_{2}\sum_{[\mathbf{i},\mathbf{j}]}\widehat{\mathbf{S}}_{\mathbf{i}}
\widehat{\mathbf{S}}_{\mathbf{j}}+
J_{3}\sum_{\{\mathbf{i},\mathbf{j}\}}\widehat{\mathbf{S}}_{\mathbf{i}}
\widehat{\mathbf{S}}_{\mathbf{j}}
\label{Hamilt}
\end{equation}
where $(\widehat{\mathbf{S}}_{\mathbf{i}})^2=3/4$,
$\langle\mathbf{i},\mathbf{j}\rangle$ denotes NN (nearest neighbor) bonds,
$[\mathbf{i},\mathbf{j}]$ denotes NNN (next-nearest neighbor)
bonds and $\{\mathbf{i},\mathbf{j}\}$ denotes
NNNN (next-to-next-nearest neighbor) bonds of the square lattice
sites $\mathbf{i},\mathbf{j}$.

Expression \ref{Hamilt} provides the minimal possible model, since quantum (and classical in the limit $S \to \infty$) helices appear starting from ``$J_3$''-level of multi-exchange Heisenberg Hamiltonian. In other words,
$J_1-J_2$ yet does not lead to a helical state.

We first briefly remind the classical limit of the problem. For classical spins in 2D any order, commensurate or incommensurate, can be set by the simple ansatz (plane spiral),~\cite{Luttin46_PR,Villai77_JdP,Misgui05_FruSpin} see also symmetry analysis in Ref.~\cite{Messio11_PRB}.
\begin{equation}
\mathbf{S_r} = \mathbf{e_1}\cos(\mathbf{q}_0 \mathbf{r}) + \mathbf{e_2}\sin(\mathbf{q}_0 \mathbf{r}), \label{S_r}
\end{equation}
where $\mathbf{e_1}$ and $\mathbf{e_2}$ are in-plane orthogonal unit vectors.
For fixed values of exchanges  $J_1, J_2, J_3$ the spin structure is determined by the energy minimization with respect to control point $\mathbf{q}_0$ position.

First of all this means that only long-range order (LRO) is realized in the classical limit, no short-range order (SRO), that is no spin liquid.
The relation \ref{S_r} means $\delta$-function-like spin-spin correlation functions.

In the quantum case under consideration ($S = 1/2$), we underline, average site spin is zero
\begin{equation}
\langle \mathbf{S_r} \rangle = 0,  \label{S_0}
\end{equation}
and the spin order is defined by the structure factor which is usually a complicated continuous function of momentum $\mathbf{q}$ in the Brillouin zone with more or less pronounced maximum.

\subsection{The method}\label{sec.method}

We use the so called spherically symmetric self-consistent approach for spin-spin Green's functions (SSSA).~\cite{Kondo72_PTP,Shimah91_JPSJ,Baraba94_JPSJ,Hartel13_PRB,Mikhey16_JMMM,Mikhee18_JETP}

SSSA conserves all the symmetries of the problem: the $SU(2)$-spin symmetry and the translational invariance and allows:

\textbf{i.} to hold the Marshall and Mermin–Wagner theorems (in our case it means in particular that average site spin is zero at any temperature, see Eq.~\ref{S_0}).

\textbf{ii.} to analyse at $T=0$ the states with and without long-range order

\textbf{iii.} to find in the wide temperature range: the spin-excitation spectrum $\omega_{\mathbf{q}}$, the dynamic susceptibility $\chi(\mathbf{q},\omega,T)$ and the structure factor $c_{\mathbf{q}}$.

Note that SSSA always leads to a singlet state. At $T=0$ and under Marshall’s theorem conditions the approach does not contradict the theorem. At $T > 0$ and arbitrary exchange couplings $J_1$, $J_2$, $J_3$ the approach is consistent with Mermin-Wagner theorem.
Note also we don't know any descriptions of broken-symmetry states, e.g., box or columnar spin liquid states as well as fractionalized excitations by SSSA or related approaches.~\cite{Hermel05_PRB}

The core of the SSSA is comprised by the chain of equations for spin Green's function
\begin{equation}
G^{zz}_{\mathbf{nm}}=
\langle S_{\mathbf{n}}^{z}|S_{\mathbf{m}}^{z}\rangle
_{\omega +i\delta }=-i\int\limits_{0}^{\infty }dt\,e^{i\omega t}\langle
[ S_{\mathbf{n}}^{z}(t),S_{\mathbf{m}}^{z}]\rangle
\end{equation}
truncated at the second step.

The spherical symmetry is maintained
$G^{\alpha \beta}_{\mathbf{nm}} \propto \delta_{\alpha \beta}$, $\alpha, \beta = x,y,z$, average cite spin is zero $\langle S_{\mathbf{n}}^{\alpha }\rangle = 0$,
three branches of spin excitations are degenerate with respect to $\alpha $.
The spin order (short- or long-range) is characterized by spin--spin correlation functions. The long-range order possible only for $T = 0$ is featured by spin--spin correlation non-vanishing at infinity. Hereafter we focus on $T \neq 0$.

The $(\mathbf{q},\omega)$-dependent Green's function
\begin{equation}G(\mathbf{q},\omega,T)=\langle
S_{\mathbf{q}}^{z}|S_{-\mathbf{q}}^{z}\rangle _{\omega }, \quad S_{\mathbf{q}}^{z}=\frac{1}{\sqrt{N}}{\,}\sum\limits_{\mathbf{r}}
e^{-i\mathbf{qr}}S_{\mathbf{r}}^{z} ,
\end{equation}
acquires the form
\begin{equation}
G(\mathbf{q},\omega,T)=\frac{F_{\mathbf{q}}}{(\omega+i0) ^{2}-\omega
_{\mathbf{q}}^{2}}  \label{a_GFmf1},
\end{equation}
see Ref.~\cite{Baraba11_TMP} for supplementary details and bulky expressions for $T$-depending $F_{\mathbf{q}}$ and the spin excitations spectrum $\omega_{\mathbf{q}}$. Here the damping of spin excitation is neglected (``quasielastic'' approximation).

In (\ref{a_GFmf1}) the numerator $F_{\mathbf{q}}$ and spin excitation spectrum $\omega _{\mathbf{q}}$ are
\begin{equation}
F_{\mathbf{q}}=8\sum_{i\in \{1,2,3\}}J_{r}(\gamma _{i}-1)c_{|\mathbf{r_i}|};
\label{JJJ_Fq}
\end{equation}
\begin{equation}
\omega _{\mathbf{q}}^{2}=2\sum_{\alpha=1}^{12}\Gamma _{\alpha}K_{\alpha}.
\label{JJJ_Ome2_q}
\end{equation}

Lattice sums in (\ref{JJJ_Ome2_q}) $K_{\alpha}$ have the form:
\begin{eqnarray*}
K_{\alpha=1}&=& J_{1}J_{2}\mathcal K_{1,2}+J_{1}J_{3}\mathcal K_{1,3}+ \nonumber \\
&&+J_{1}^{2}(z_{1}(z_{1}-1)\tilde{c}_{|\mathbf{r_1}|}+z_{1}c_{|\mathbf{r}|=0}+\mathcal K_{1,1});  \nonumber \\
K_{\alpha=2}&=& J_{2}J_{1}\mathcal K_{2,1}+J_{2}J_{3}\mathcal K_{2,3}+ \nonumber \\
&&+J_{2}^{2}(z_{2}(z_{2}-1)\tilde{c}_{|\mathbf{r_2}|}+z_{2}c_{|\mathbf{r}|=0}+\mathcal K_{2,2});  \nonumber \\
K_{\alpha=3}&=&-J_{1}^{2}z_{1}^{2}\tilde{c}_{|\mathbf{r_1}|};\quad \quad
K_{\alpha=4}=-J_{2}^{2}z_{2}^{2}\tilde{c}_{|\mathbf{r_2}|};  \nonumber \\
K_{\alpha=5}&=&-J_{1}J_{2}z_{1}z_{2}\tilde{c}_{|\mathbf{r_1}|};\quad K_{\alpha=6}=-J_{1}J_{2}z_{1}z_{2}
\tilde{c}_{|\mathbf{r_2}|};  \nonumber \\
K_{\alpha=7}&=& J_{3}J_{1}\mathcal K_{3,1}+J_{3}J_{2}\mathcal K_{3,2}+ \nonumber \\
&&+J_{3}^{2}(z_{3}(z_{3}-1)%
\tilde{c}_{|\mathbf{r_3}|}+z_{3}c_{|\mathbf{r}|=0}+\mathcal K_{3,3});  \nonumber \\
K_{\alpha=8}&=&-J_{3}^{2}z_{3}^{2}\tilde{c}_{|\mathbf{r_3}|};\quad \quad
K_{\alpha=9}=-J_{1}J_{3}z_{1}z_{3}\tilde{c}_{|\mathbf{r_1}|};  \nonumber \\
K_{\alpha=10}&=&-J_{3}J_{1}z_{3}z_{1}\tilde{c}_{|\mathbf{r_2}|};\quad
K_{\alpha=11}=-J_{2}J_{3}z_{2}z_{3}\tilde{c}_{|\mathbf{r_2}|};  \nonumber \\
K_{\alpha=12}&=&-J_{3}J_{2}z_{3}z_{2}\tilde{c}_{|\mathbf{r_3}|};  \label{JJJ_Ks}
\end{eqnarray*} here
\begin{equation}
\mathcal K_{i,j}=\sideset{_{}^{}}{^{'}} \sum_{\mathbf{n}_{i},\mathbf{n}_{j}}
\tilde{c}_{|\mathbf{r}_{i}+\mathbf{r}_{j}|}, \label{JJJ_KK}
\end{equation}
where $\mathbf{r}_{i}$, $\mathbf{r}_{j}$ are radius vectors of nearest, next-nearest or next-to-next-nearest neighbor sites; $\mathbf{n}_{i} = \mathbf{r}_{i}/|\mathbf{r}_{i}|$, and $\sum'$ implies $\mathbf{r}_{i}\ne \mathbf{r}_{j}$.

The coefficients $\Gamma _{i}$ in the expression for $\omega _{\mathbf{q}}$ are:
\begin{eqnarray}
\Gamma_{1}&=& 1-\gamma_{1};\quad \quad \quad \Gamma_{2}=1-\gamma_{2};
\nonumber \\
\Gamma_{3}&=& 1-\gamma_{1}^{2};\quad \quad \quad \Gamma_{4}=1-\gamma_{2}^{2};
\nonumber \\
\Gamma_{5}&=& (1-\gamma_{1})\gamma_{2};\quad \
\Gamma_{6}=(1-\gamma_{2})\gamma_{1};  \nonumber \\
\Gamma_{7}&=& 1-\gamma_{3};\quad \quad \ \ \Gamma_{8}=1-\gamma_{3}^{2};
\nonumber \\
\Gamma_{9}&=& (1-\gamma_{1})\gamma_{3};\quad
\Gamma_{10}=(1-\gamma_{3})\gamma_{1};  \nonumber \\
\Gamma_{11}&=& (1-\gamma_{2})\gamma_{3};\quad
\Gamma_{12}=(1-\gamma_{3})\gamma_{2}.  \label{JJJ_Gams}
\end{eqnarray}

In (\ref{JJJ_Fq}), (\ref{JJJ_Gams})
$\gamma _{i}=(1/{z_{i}})\sum_{\mathbf n_i}e^{i\mathbf{q n}_i r_i}$, where
the sum is take over the cites of the $i$-th coordination sphere and $z_i$ is the number that cites. For 2D square lattice $z_{1}=z_{2}=z_{3}=4$.
In eq.~(\ref{JJJ_KK}), $\tilde{c}_{|\mathbf{r}_i|}$ are correlators ${c}_{|\mathbf{r}_i|}$ with vertex corrections; we use here the one vertex approximation (see, e.g.,~\cite{Baraba11_TMP},\cite{Baraba03_TaFra}).

In other words, SSSA truncates the equation-of-motion hierarchy for the
spin-spin Green's function after the second level, yielding ultimately lorentzian spin excitations, eq.~\ref{a_GFmf1}, with self-consistently determined pole weights $F_{\mathbf{q}}$ and positions $\omega_{\mathbf{q}}$.

For $J_{1}-J_{2}-J_{3}$ model the Green's function $G(\mathbf{q},\omega,T)
$ depend on the correlators $c_{\mathbf{r}} = c_{|\mathbf{r}|}=\langle S_{\mathbf{n}}^{z}S_{\mathbf{n}+\mathbf{r}}^{z}\rangle $
for eight coordination spheres (we use the definition the first coordination sphere as the manifold of the nearest neighbors, the second – as the manifold of next nearest neighbors and so on).
Moreover, $G(\mathbf{q},\omega,T)$ must satisfy the spin constraint, the on-site correlator $c_{\mathbf{r=0}}=\left\langle
S_{\mathbf{n}}^{z}S_{\mathbf{n}}^{z}\right\rangle =1/4$.
All the correlators can be evaluated self-consistently in terms of $G(\mathbf{q},\omega,T)$.
So there are nine conditions
\begin{equation}
c_{\mathbf{r_k}}=\frac{1}{N}\sum_{\mathbf{q}}c_{\mathbf{q}}e^{i\mathbf{qr_k}}; \label{a_Cr_1}
\end{equation}
where $\mathbf{r_0} = 0$, $\mathbf{r_i}$ $(i = 1 ,\ldots, 8$) belongs to $i$-th coordination spheres, the structure factor $c_{\mathbf{q}}$
\begin{equation}
c_{\mathbf{q}}=\left\langle
S_{\mathbf{q}}^{z}S_{\mathbf{-q}}^{z}\right\rangle =
-\frac{1}{\pi}
\int_{0}^{\infty }d\omega \coth\left(\frac\omega{2T}\right) \mathrm{Im}\,G(\mathbf{q},\omega,T).
\label{a_Cq_1}
\end{equation}

The system of self-consistent equations \ref{a_GFmf1}--\ref{a_Cq_1}
is analyzed numerically. Hereafter all the energy-related parameters are set in the units of $J=\sqrt{J_{1}^{2}+J_{2}^{2}+J_{3}^{2}}$.

The structure factor landscape saturates at $T \ll 1$. So all the foregoing results have been obtained at low temperature $T = 0.02$.

Indeed, Green's-function method has its limitations. In the case of better investigated parent J1-J2 square-lattice model SSSA in its standard simple realization is known to overestimate the borders of the spin-liquid region  (see Fig.1 from \cite{Siurak01_PRB}) compared to results of exact diagonalization, density-matrix renormalization group~\cite{Jiang12_PRB}, and functional renormalization group~\cite{Reuthe11_PRB}.
Here we use the variant of SSSA, where we neglect the spin-spin excitations damping and involve only one vertex correction. Note that in a number of previous works this limitation was partially eliminated and fine tuning of the approach was elaborated including damping, many vertices approximation, Zwanzig–Mori projection approach~\cite{Mikhey16_JMMM,Baraba11_TMP,Baraba15_JL}. It was shown in particular that under the tuning of the SSSA spin-liquid area is close to that of modern numerical methods. Nevertheless, the quantitative picture remains stable.

\subsection{Results and discussion}\label{sec.resdis}
\begin{figure}[tb]
  \centering
  \includegraphics[width=.5\columnwidth]{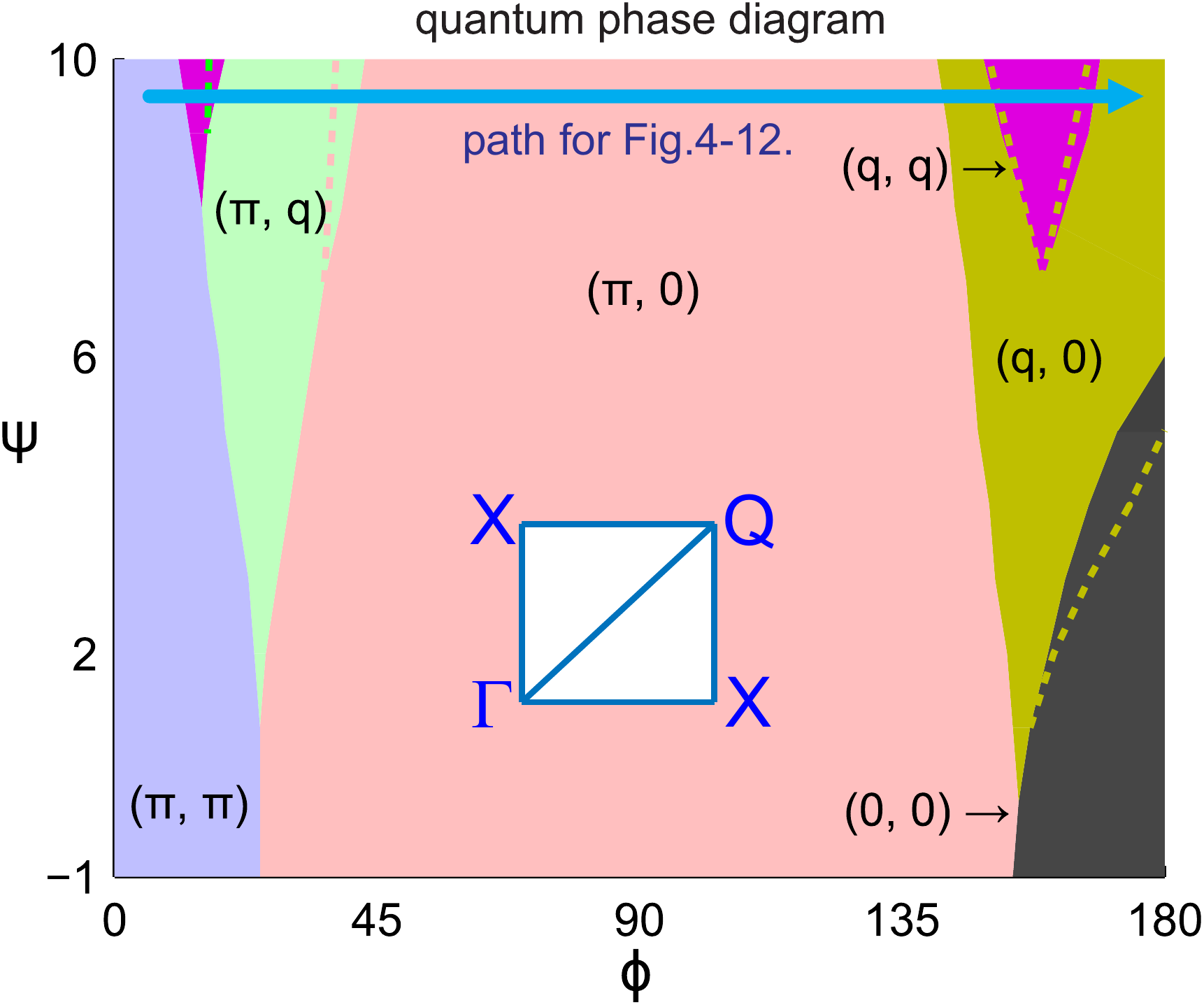}\\
  \caption{(Color online) ``Phase diagram'' for the problem at hand. Different colors correspond to different spin local order. The labels mark the positions of structure factor maxima. The evolution of structure factor in figures below mainly follows the thick blue arrow line.
  Exchange integrals are parameterized as in Fig.~\ref{Globe}. Solid borders correspond to temperature $T=0.4$, dashed ones --- to $T=0.2$. At lower temperatures local order boundaries stabilize. We note that though Fig.~3 and Fig.~1 are similar, the physical phenomena behind them are quite different.
  The inset shows quarter of the Brillouin zone, where ${\mathbf{\Gamma}}= (0,0)$, $\mathbf{X}=(\pi,0),(0,\pi)$ and $\mathbf{Q}=(\pi,\pi)$.
  }\label{Phases}
\end{figure}
In the classical limit the structure factor is always $\delta$-function like (see Eq.~\ref{S_r}). This means that there is only one unique wave vector $\mathbf{q}$ defining spin order (apart from symmetry equivalent points in the Brillouin zone).

In the quantum case $S=1/2$ the structure factor is usually a smooth complicated continuous function of momentum $\mathbf{q}$. Nevertheless at not very high temperatures local spin order can be distinguished by the positions of the structure factor maxima (see Fig.~\ref{Phases}).

The most interesting situation corresponds to continuous degeneracy of the structure factor maxima: in this case they merge into the curve in the $\mathbf{q}$-space (this is hardly possible in the classical limit). Sometimes this curve is topologically equivalent to a circle, then we can discuss the circular quantum structures.

We underline, that the last picture is natural only for the strongly frustrated model. For example such continuous degeneracy does not appear in the $J_1-J_2$ square lattice model: the third frustrating exchange $J_3$ is necessary.

Note that the adopted quantum approach treats the spin liquid rather coarsely, being based generally on two-site spin-spin correlators. There are numerous works, where at zero and extremely low temperatures phases of a more complex structure (e.g. columnar phase, spin nematic, vortex crystal, valence-bond crystal,…) are mentioned, determined by higher-order correlators (usually four-cite).
The nomenclature of such states is extensive, and the related literature is vast (see e.g.~\cite{_Diep_FrustrSpin_13,Schmid17_PR} for the review).
Here we detect a noticeable maximum of the structural factor (and the corresponding minimum of the spin gap). Against this background, correlators of higher orders will only lead to the appearance of small ripples on the main structure. Moreover, as the temperature rises, the ``fine structure'' blurs much faster than the main peaks.

Basically, it is possible to get more complex structure (columnar, nematic, etc) in the SSSA-like approach. Then one should start from the states not of a single cite, but of a block like it was done, e.g. in Ref.~\cite{Baraba89_JPCM}. This is the subject for the forthcoming investigations.

\subsubsection{Phase diagram: general properties}
\label{Phase_diagram}

In $J_1-J_2-J_3$ model the norm $\sqrt{(J_1^2+J_2^2+J_3^2)}$ is irrelevant for short-range order and the phase diagram. So the kind of ``globe'' parametrisation is convenient

\begin{eqnarray} \nonumber
  J_1 &=& \cos(\psi)\cos(\phi),
  \\
  J_2 &=& \cos(\psi)\sin(\phi),
  \\ \nonumber
  J_3 &=& \sin(\psi).
\end{eqnarray}

Here $\psi = \pi/2 - \theta$, and $\theta$ is the standard spherical angle. This choice improves the observables readability.

The physics of the problem depends only on the relative magnitude of the coupling constants, so energy scale may be freely chosen. The standard choice in the field is $|J_1|$. Nevertheless while passing the areas $|J_1| \ll |J_2|,|J_3|$ the euclidean norm of the vector $(J_1, J_2, J_3)$ is more convenient.

Like on the earth globe, there is a ``no man's land'' at the ``poles'' ($\psi = \pm\pi/2$, that is $J_1=J_2=0$, $J_3 = \pm 1$), where there is almost nothing interesting and experimentally relevant on the phase diagram. The most intriguing are the ``equatorial'' latitudes of the ``north'' hemisphere, $0 \leq \phi \leq 2\pi$, $-\pi/2 \leq \psi \leq \pi/2$. One can see, that this region, depicted in Fig.~\ref{Phases}, is the most frustrated.

We choose the trajectory on the phase diagram, see thick blue arrow line ($J_3 = 0.17$, that is $\psi = 10^{\circ}$) in Fig.~\ref{Phases}, that passes the following states:
\begin{itemize}
  \item AFM with structure factor maxima at $\mathbf{q_0}= (\pm\pi,\pm\pi)$;
  \item stripe --- $\mathbf{q_0}= (\pm\pi,0),\, (0,\pm\pi)$, in the classical limit it would be alternating stripes along a lattice with spins up and down;
  \item FM $\mathbf{q_0}= (0,0)$;
  \item helicoid $\mathbf{q_0}= (\pm q,0),\, (0, \pm q)$;
  \item helicoid $\mathbf{q_0}= (\pm \pi,q),\, (q, \pm \pi)$;
  \item helicoid $\mathbf{q_0}= (\pm q, \pm q)$.
\end{itemize}
The last three in the classical limit would be spin helices rotating along one of the axis or along the diagonal of the square lattice. In Fig.~\ref{Phases} and hereafter we label the local orders with one of the equivalent points $\mathbf{q_0}$.

Below evolution of the structure factor and the spin excitations spectra along the trajectory is investigated. We focus on the transitions corresponding to local order changes. The borders of different local orders are well-defined and correspond to changes of structure factor maxima.

The situation in the ``depth'' of each phase is more or less clear, at least qualitatively. But the transitions between definite spin-liquid local orders is much more intriguing. Note that the physical picture here is some sense similar to liquid-liquid transitions.~\cite{Kataya00_N,Brazhk02_book,Ryltse13_PRE,Ryltse17_SM,Mitrea18_NC,Wouter18_S}

We are to remind some general properties of the spectrum.~\cite{Shimah91_JPSJ,Baraba94_JPSJ,Hartel13_PRB,Mikhey16_JMMM} The spin gap is always closed at trivial point $\mathbf{q_0}= (0,0)$ at any temperature. At $T = 0$ it might be closed at nontrivial points in the Brilloin zone with $\delta$-peak of structure factor at the same point. These means the corresponding long-range order (AFM, FM, stripe or helical). At $T = 0$ spin-liquid states are also possible.

We are interested in the case of $T > 0$, when the long-range order is always absent, but the short-range order remains pronounced and complicated. The local order is defined by the structure factor maximum and the spectrum minimum at nontrivial points.

\subsubsection{From AFM via two helices to stripe}

\paragraph{From $(\pi,\pi)$ via $(q,q)$ to $(\pi,q)$.}
\begin{figure}
  \centering
  \includegraphics[width=\textwidth]{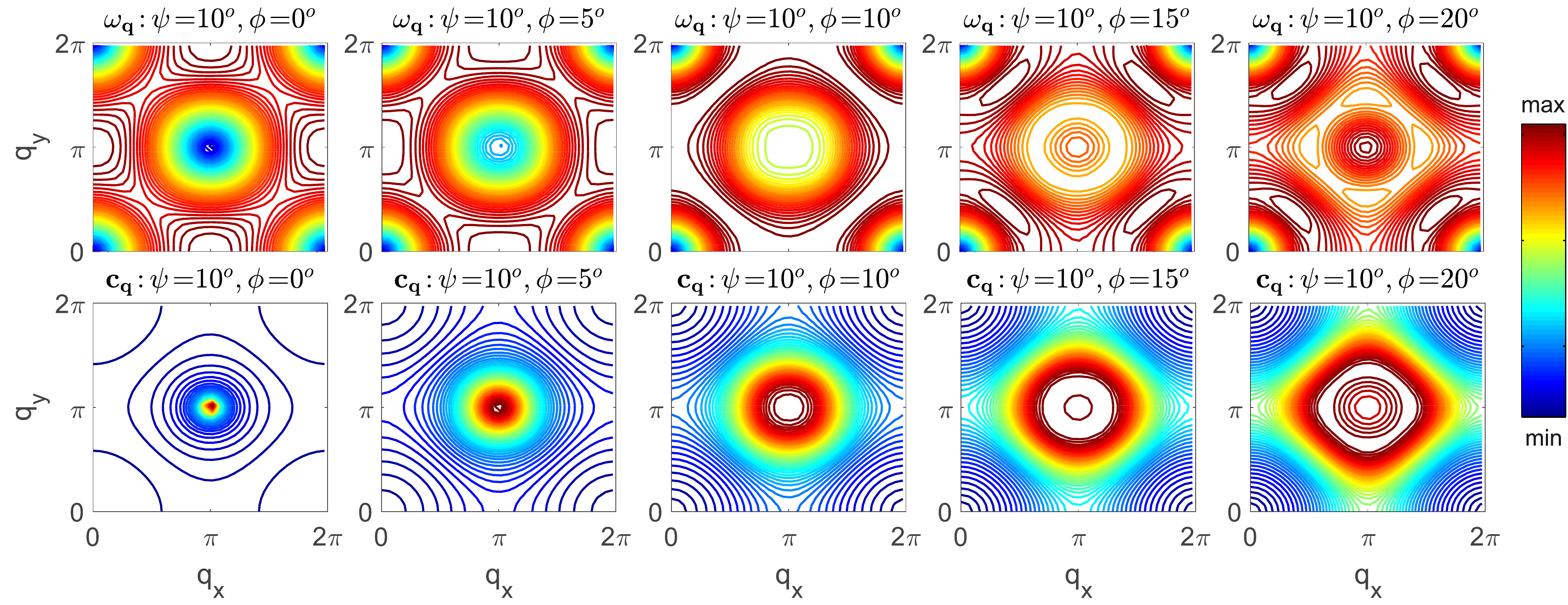}\\
  \caption{(Color online) Contour lines for excitation spectra $\omega_{\mathbf{q}}$ (upper row) and the structure factor $c_{\mathbf{q}}$ (lower row). Exchanges $J_{1}$, $J_{2}$ and $J_{3}$ are parameterized by spherical angles $\phi$ and $\psi$ (in degrees):
  $J_1 = \cos \psi \cos \phi$, $J_2 = \cos \psi \sin \phi$,
  $J_3 = \sin \psi$. Here $\psi=10^{\circ}$ and
  $\phi \in [ 0^{\circ}, 20^{\circ}]$.
  On the first column $\omega_{\mathbf{q}}$ minimum and $c_{\mathbf{q}}$ maximum at AFM point $(\pi, \pi)$ indicate AFM short-range order. With growing $\phi $ AFM gap is opening and circular $c_{\mathbf{q}}$ structure is developed, acquiring then square features.
  }\label{phi0_to_phi20}
\end{figure}
\begin{figure}[t]
  \includegraphics[width=0.75\columnwidth]{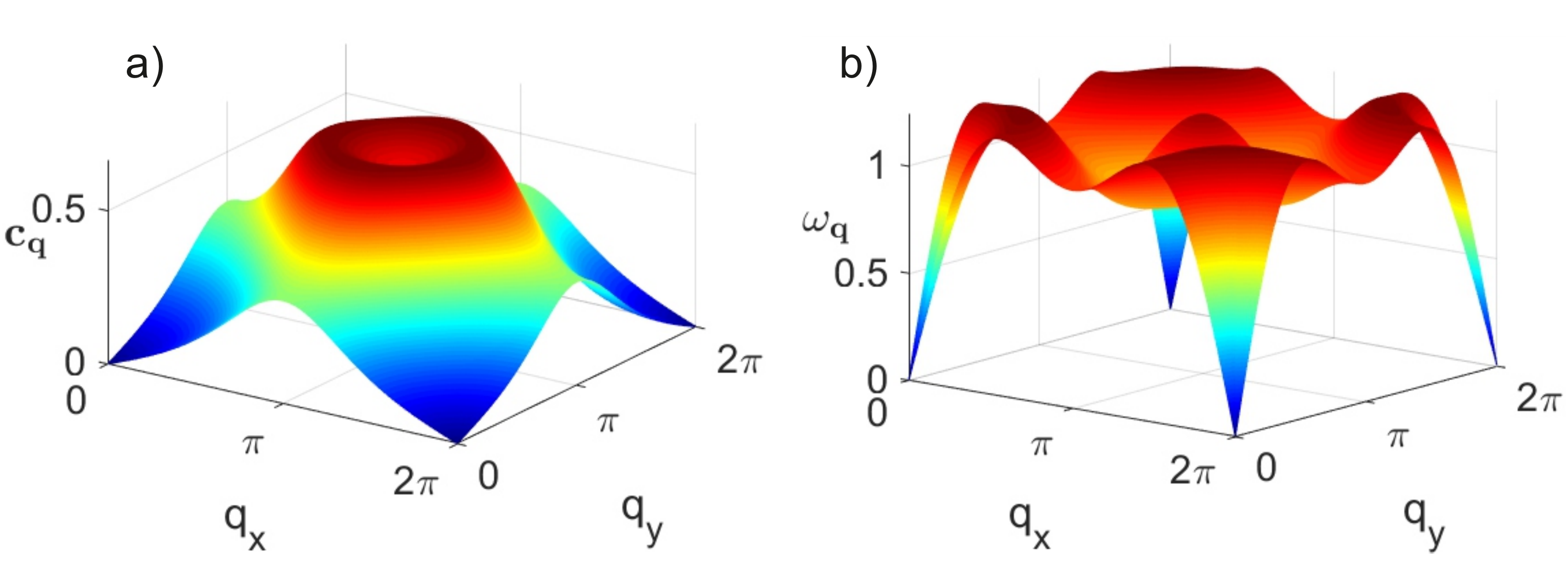}
  \caption{(Color online) The a) ``volcanic'' and b) ``spider'' figures show structure factor $c_{\mathbf{q}}$ and the spin excitations spectrum $\omega_{\mathbf{q}}$. Here $\psi = 10^{\circ}$, $\phi = 20^{\circ}$ that correspond to the region of AFM circular states.
  }\label{gyrate_AFM}
\end{figure}
The spectrum and structure factor evolution in this domain is shown in Fig.~\ref{phi0_to_phi20}. We have chosen the frame of reference for the Brillouin zone $(0 \leq q_{x,y} \leq 2\pi)$. In this case the AFM maxima are located in the centre of the Brillouin zone.

\begin{figure*}[t]
  \centering
  \includegraphics[width=\textwidth]{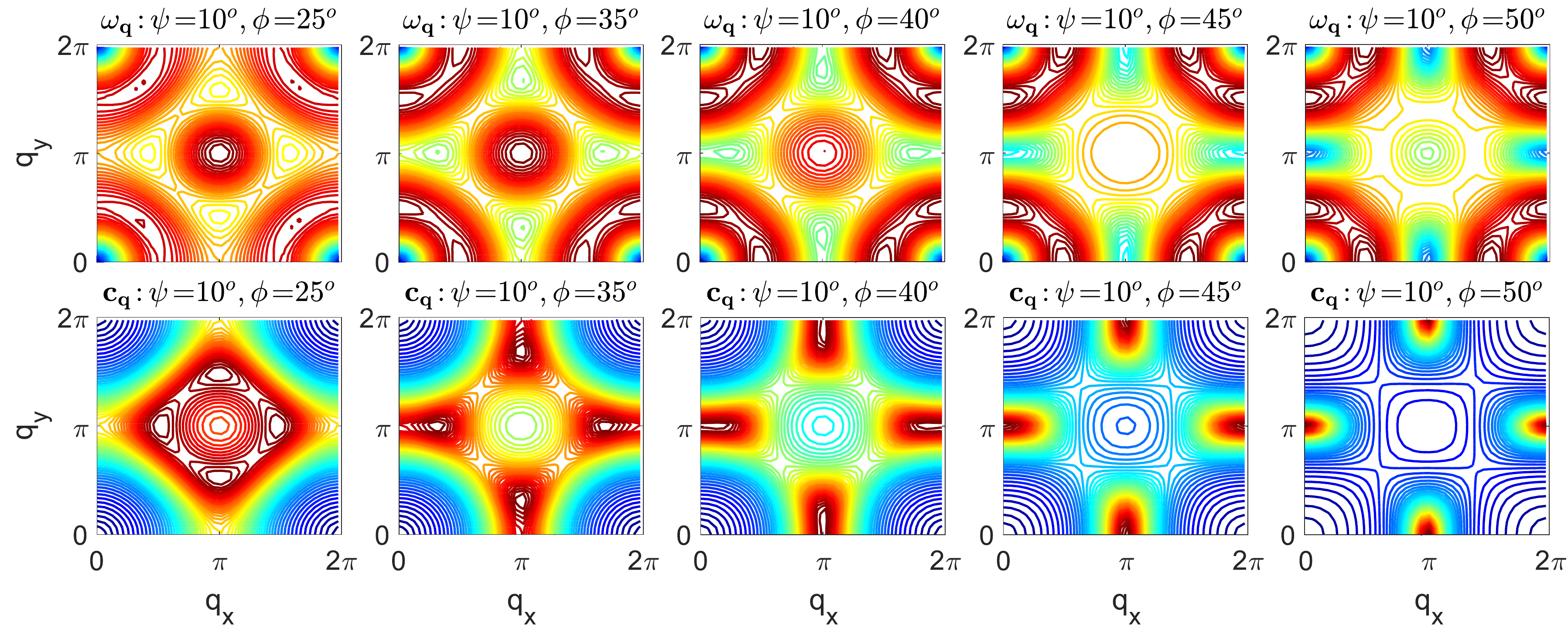}\\
  \caption{(Color online) The same as in Fig.~\ref{phi0_to_phi20} (contour lines for $\omega_{\mathbf{q}}$ and $c_{\mathbf{q}}$), for
  $\psi=10^{\circ}$ but $\phi \in [25^{\circ}, 50^{\circ}]$.
  Here local order is evolving from complex $(\pi, q)$ helix with $c_{\mathbf{q}}$ maxima forming the modulated square line to stripe order with $c_{\mathbf{q}}$ maximum at $(\pi, 0)$, see also Fig.~\ref{prestripe}.
  }\label{phi25_to_phi50}
\end{figure*}

The first figure-column in Fig.~\ref{phi0_to_phi20} just corresponds to AFM with the sharp maximum of the structure factor $c_{\mathbf{q}}$ and the local minimum of the spin excitations spectrum $\omega_{\mathbf{q}}$ at the AFM point $(\pi,\pi)$.
For large enough $\phi$ ($\phi \gtrsim 40^{\circ}$) the short-range order becomes clearly stripe-like (see Fig.~\ref{phi25_to_phi50}) with $c_{\mathbf{q}}$ maximum and $\omega_{\mathbf{q}}$ minimum at the stripe point $(\pi,0)$ and the equivalent ones.
The half-width of the mentioned maxima in these limits defines the correlation length correspondingly for AFM and stripe order.

In between these limits $c_{\mathbf{q}}$ evolves smoothly and its peak becomes much wider implying the correlation length's diminishing, see the second figure-column in Fig.~\ref{phi0_to_phi20}.

\begin{figure}[t]
\centering
  \includegraphics[width=0.6\columnwidth]{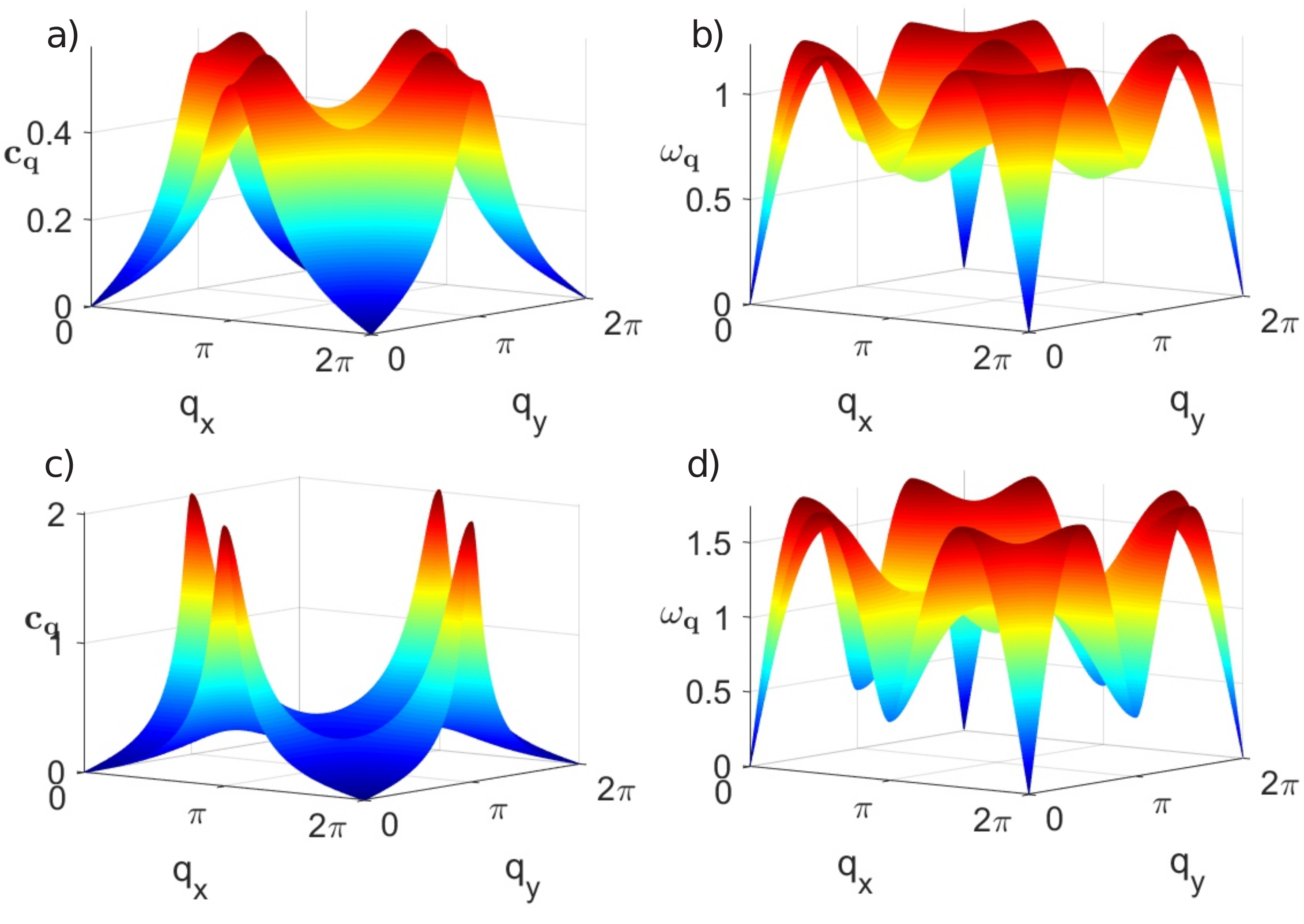}\\
  \caption{(Color online) The evolution of structure factor $c_{\mathbf{q}}$ and the spin excitations spectrum $\omega_{\mathbf{q}}$ from $(\pi,q)$ helical local order (top row,  $\psi = 10^{\circ}$, $\phi = 35^{\circ}$) to  stripe $(\pi,0)$ one (bottom row,  $\psi = 10^{\circ}$, $\phi = 50^{\circ}$).
  In the first case (a) the ``volcanic'' structure of $c_{\mathbf{q}}$ can be still traced, but the circular degeneracy maxima manifold has disappeared.
  In the second (c) case $c_{\mathbf{q}}$ is already stripe-like.
  In terms of $\omega_{\mathbf{q}}$ this transformation, from (b) to (d), looks like the growth of additional four ``legs'' of the spider-spectrum.
  }\label{prestripe}
\end{figure}
At higher $\phi$ (that is $J_2$) the top of the $c_{\mathbf{q}}$ peak starts collapsing down and the peak acquires ``volcanic'' shape, see the evolution between second and fifth figure-columns in Fig.~\ref{phi0_to_phi20}.

\begin{figure}[t]
  \centering
  \includegraphics[width=\textwidth]{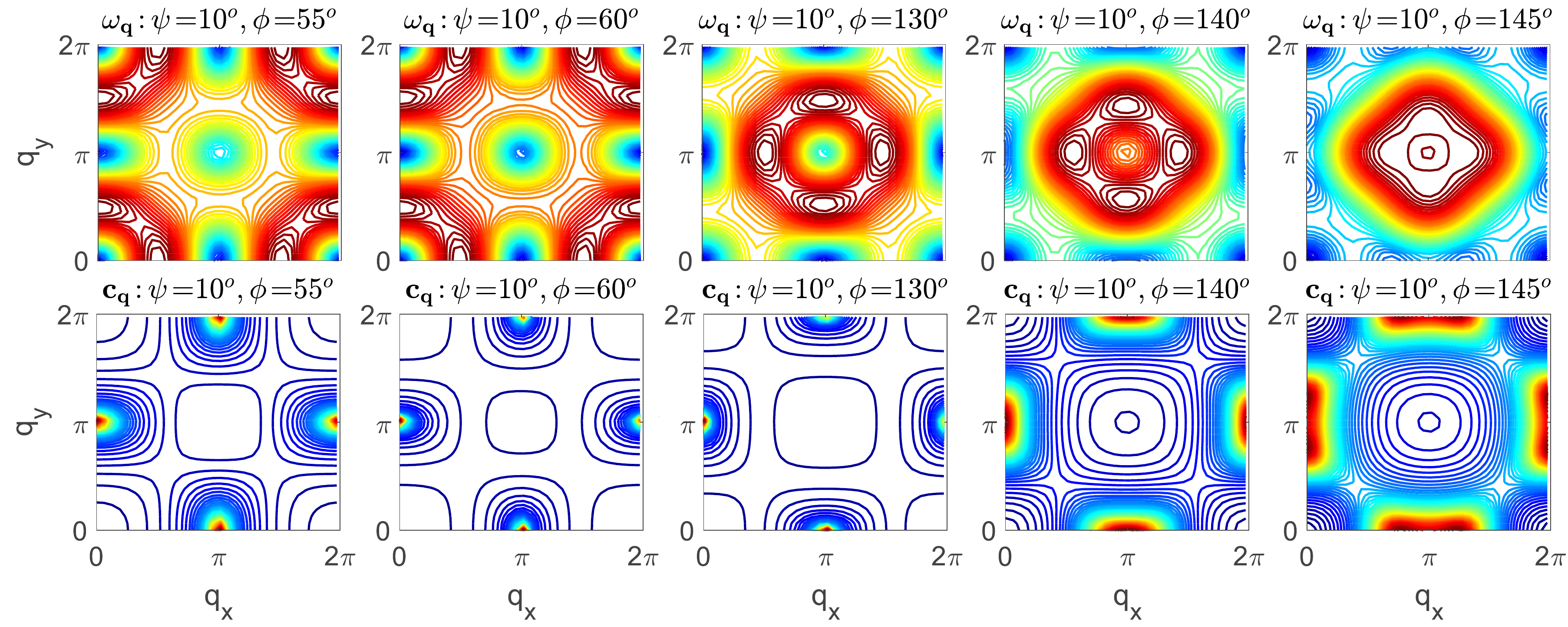}\\
  \caption{(Color online) The same as in Fig.~\ref{phi0_to_phi20} (contour lines for $\omega_{\mathbf{q}}$ and $c_{\mathbf{q}}$), for
  $\psi=10^{\circ}$ but $\phi \in [55^{\circ}, 145^{\circ}]$.
  The first three figure-columns represent the stripe-state with  $\omega_{\mathbf{q}}$ minimum and $c_{\mathbf{q}}$ maximum at point $(\pi,0)$ (and at equivalent points). We remind that the correlation length is related
  to the width of $c_{\mathbf{q}}$ maximum. The correlation length diminishes from left to right. The last two figure-columns correspond to continuous splitting of stripe $c_{\mathbf{q}}$ maximum, that can be interpreted as the crossover to the $(q,0)$ incommensurate helical state, more exactly, to the quantum superposition of several such states. See also Fig.~\ref{after_stripe}.
  }
  \label{phi55_to_phi145}
\end{figure}
\begin{figure}[t]
\centering
  \includegraphics[width=0.6\columnwidth]{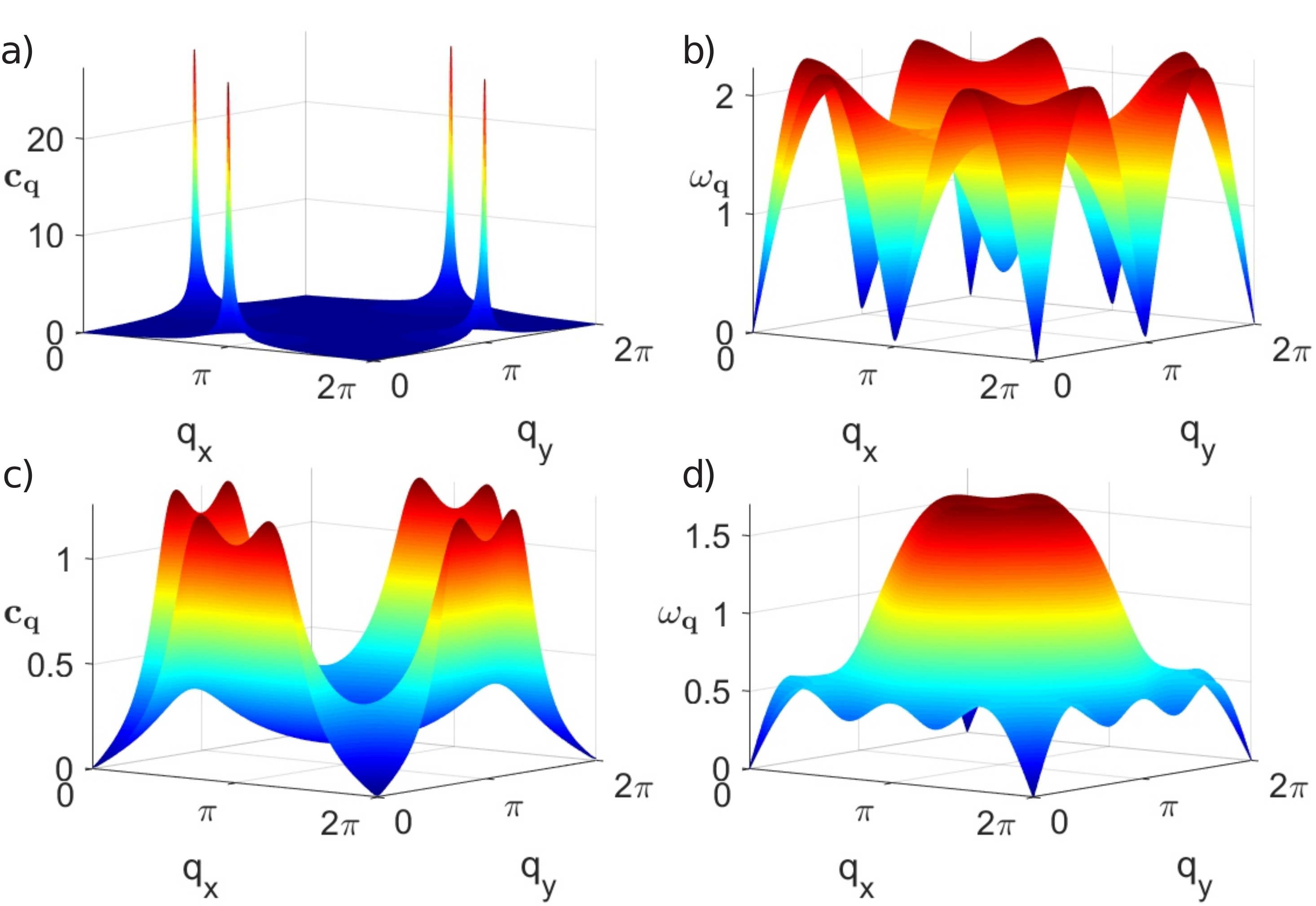}\\
  \caption{(Color online) The evolution of structure factor $c_{\mathbf{q}}$ and the spin excitations spectrum $\omega_{\mathbf{q}}$ from stripe $(\pi,0)$ local order (top row,  $\psi = 10^{\circ}$, $\phi = 60^{\circ}$) to helical $(q,0)$ one (bottom row,  $\psi = 10^{\circ}$, $\phi = 145^{\circ}$).
  Note strong difference in z-scales for $c_{\mathbf{q}}$. The growth of $\phi$ induces the split of the $c_{\mathbf{q}}$ peaks. In terms of $\omega_{\mathbf{q}}$ this transformation, from (b) to (d), looks like the growth of additional four ``legs'' of the spectrum (transformation from ``spider'' to ``squid'' shape).
  }\label{after_stripe}
\end{figure}

The form of the structure factor defines the symmetry and the structure of the underlying quantum state. Thus, we get the desired circular quantum states,
with the $c_{\mathbf{q}}$ maxima forming the circle structure centered at $(\pi,\pi)$ see Fig.~\ref{gyrate_AFM}. This indicates local order of the antiferromagnetic isotropical helix~\cite{Reuthe11_PRB,Iqbal16_PRB}.
The continuous circular degeneracy can be treated as the quantum superposition of incommensurate spiral states propagating in all directions.

The $c_{\mathbf{q}}$ in Fig.~\ref{gyrate_AFM} with the volcanic shape being imaginatively squeezed to the point $(\pi,\pi)$ acquires purely AFM local order. The nonzero diameter of the $c_{\mathbf{q}}$ crater is the incommensurability parameter  for the degenerate set of helices and the width of the walls of the crater defines the correlation length.

\paragraph{From  $(\pi,q)$ to $(\pi,0)$.}

The spectrum and structure factor evolution in this domain is shown in Fig.~\ref{phi25_to_phi50}.
With the growth of $\phi$ (that is $J_2$) circular ``volcanic'' structure of $c_{\mathbf{q}}$ acquires the four-fold modulation that finally transforms into four distinct peaks. The last is the quantum stripe state: the superposition of local stripes along perpendicular directions, see Fig.~\ref{prestripe}.

In terms of spin excitations spectrum this transformation is the shift of $\omega_{\mathbf{q}}$ local minimum from the incommensurate point $(\pi,q)$ to stripe point $(\pi,0)$ with the simultaneous reduction of the corresponding spin gap, see Figs.~\ref{phi25_to_phi50}-\ref{prestripe}.

Note that the spectrum $\omega_{\mathbf{q}}$ in Figs.~\ref{phi25_to_phi50}-\ref{prestripe} has in some directions roton form, that is its local minimum in the depth of the Brillouin zone. The same is true for Fig.~\ref{gyrate_AFM}.

\subsubsection{From stripe via two helices to FM}

The spectrum and structure factor evolution in this domain is shown in Fig.~\ref{phi55_to_phi145}-\ref{after_stripe}. We remind that the frame of reference for the Brillouin zone here is $(0 \leq q_{x,y} \leq 2\pi)$.

In the range $\phi \sim 90^{\circ}\pm 30^{\circ}$ the local order is stripe-like. The correlation length is maximal for $\phi = 90^{\circ}$ and decays on both sides. After leaving the stripe region ($\phi \gtrsim 120^{\circ}$) the peaks of $c_{\mathbf{q}}$ split and the local order acquires $(q,0)$ helical structure.

Correspondingly, the excitation spectrum $\omega_{\mathbf{q}}$ undergoes the splitting of local minima and transforms from ``spider'' to ``squid'' shape.

\begin{figure}[t]
  \centering
  \includegraphics[width=\textwidth]{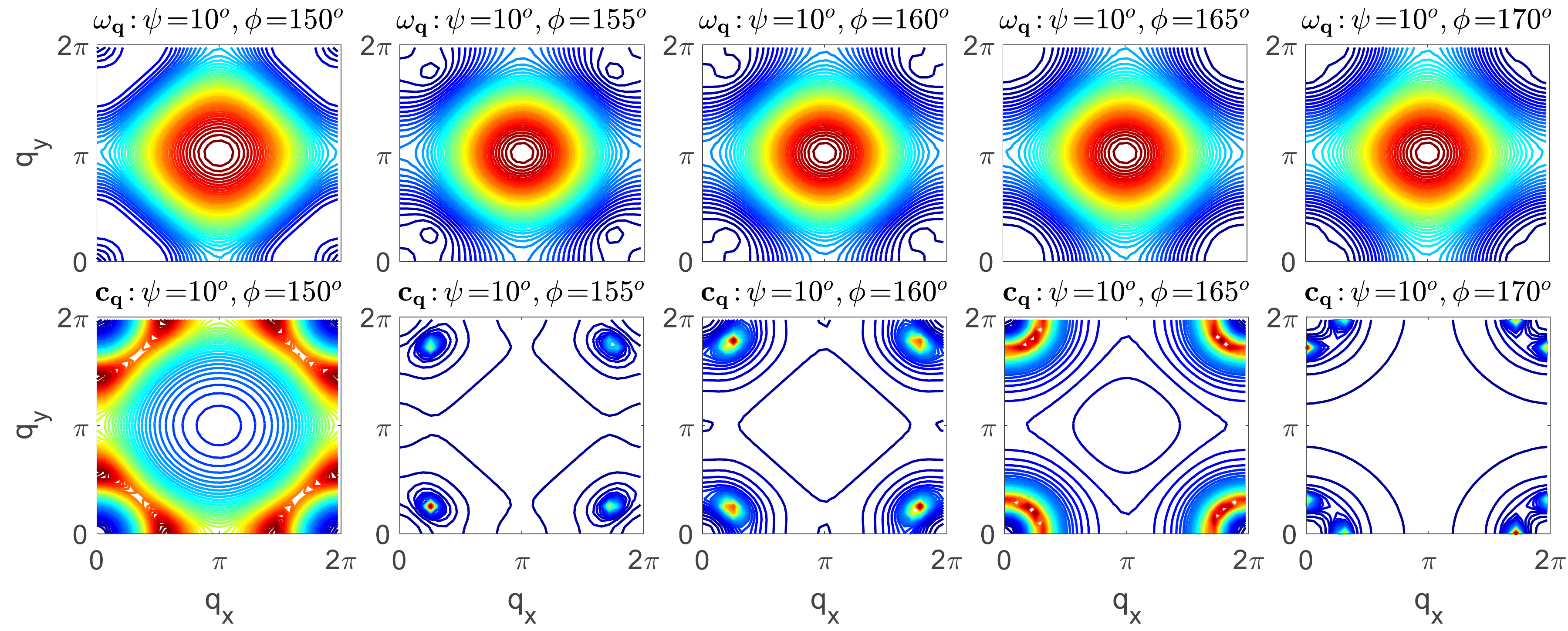}\\
  \caption{  (Color online) The same as in Fig.~\ref{phi0_to_phi20} (contour lines for $\omega_{\mathbf{q}}$ and $c_{\mathbf{q}}$), for
  $\psi=10^{\circ}$ but $\phi \in [150^{\circ}, 170^{\circ}]$.
  The first figure-column with splitted $c_{\mathbf{q}}$ maximum still corresponds to $(q,0)$ helical order. The second and third figure-columns --- $(q,q)$ helical order. And the last one shows the reentrance to $(q,0)$ order. See also Fig.~\ref{sopli}.
  }\label{phi150_to_phi170}
\end{figure}
\begin{figure}
\centering
  \includegraphics[width=0.6\columnwidth]{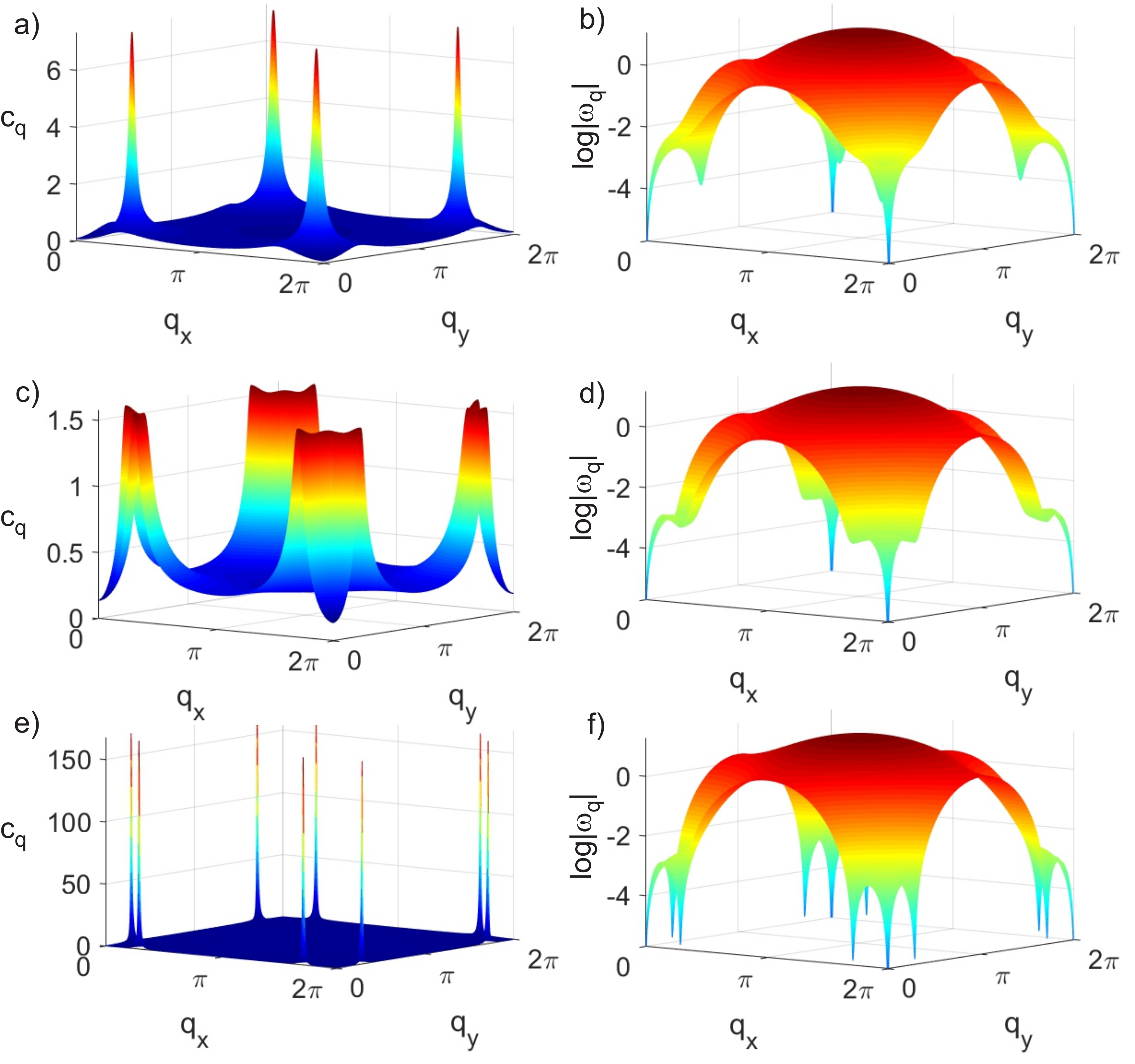}\\
  \caption{(Color online) The evolution of structure factor $c_{\mathbf{q}}$ and the spin excitations spectrum $\omega_{\mathbf{q}}$ from $(q,q)$ helical local order (top row,  $\psi = 10^{\circ}$, $\phi = 155^{\circ}$) to helical $(q,0)$ one (bottom row,  $\psi = 10^{\circ}$, $\phi = 170^{\circ}$) via FM circular state (middle row,  $\psi = 10^{\circ}$, $\phi = 160^{\circ}$).
  Note that in the usual frame of reference for the Brillouin zone $(-\pi \leq q_{x,y} \leq \pi)$ the structure factor $c_{\mathbf{q}}$ maxima form the circle line.
  }\label{sopli}
\end{figure}

Structure factor $c_{\mathbf{q}}$ maximum underdoes similar splitting, that can be interpreted as the crossover to the $(q,0)$ incommensurate helical state, more exactly, to the quantum superposition of several such states.

\paragraph{Reentrance from $(q,q)$ to $(q,0)$.}

\begin{figure}[t]
\centering
  \includegraphics[width=0.75\columnwidth]{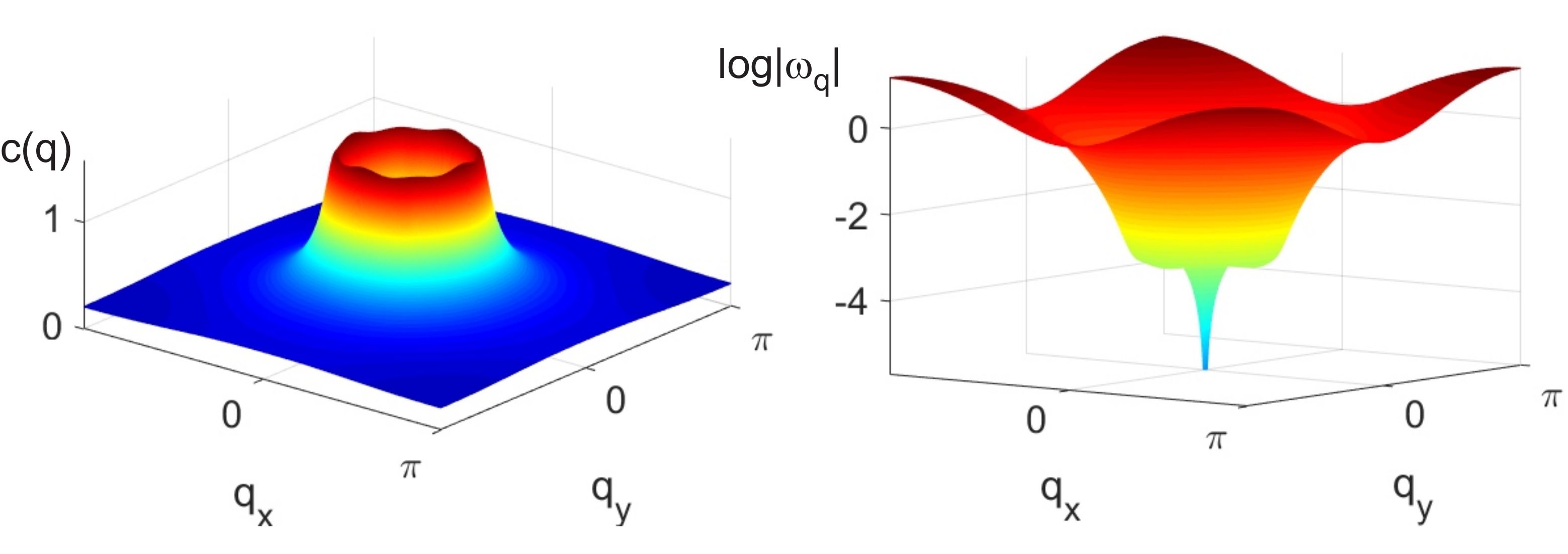}\\
  \caption{(Color online) The structure factor $c_{\mathbf{q}}$ and the spin excitations spectrum $\omega_{\mathbf{q}}$ for the same parameters like in Fig.~\ref{sopli} but now we take the usual frame of reference for the Brillouin zone: $(-\pi \leq q_{x,y} \leq \pi)$. This figure highlights that the structure factor again has circular form, however now it is FM circular state.
  }\label{kruglye}
\end{figure}

The spectrum and structure factor evolution in this domain is shown in Figs.~\ref{phi150_to_phi170}-\ref{kruglye}.
\paragraph{From $(\pi,0)$ via $(q,0)$ towards $(q,q)$.}

In contrast to the classical limit, there exists the island of $(q,q)$ helical local order with the subsequent reentrance to $(q,0)$ helical local order.

The complex helix state, that in contrast to AFM circular structure, see Fig.~\ref{gyrate_AFM}, is to be labeled as FM circular structure, appears in the borderland (see the fourth figure-column in Fig.~\ref{phi150_to_phi170}). Similar observation have been made recently in Ref.~\cite{Seabra16_PRB} using purely numerical tools (quantum Monte-Carlo simulation).

The correlation length shows nontrivial nonmonotonic evolution while passing from purely $(q,q)$ to purely $(q,0)$ helix. It dramatically drops in the borderland being sufficiently large on both sides, as it is seen from the evolution of the structure factor peaks in Fig.~\ref{sopli}.

From the first glance, it is difficult to detect a circular state from Fig.~\ref{sopli}. In the FM region returning to the standard frame of reference for the Brillouin zone $(-\pi \leq q_{x,y} \leq \pi)$ is natural. This is done in Fig.~\ref{kruglye}, where FM circular shape of the structure factor becomes obvious.

The ``flower'' spectrum on the right in Fig.~\ref{kruglye} requires additional explanation. The spectrum has two distinct parts --- the flower itself and the stem. The stem is determined by the spin gap at the trivial zero point $\mathbf{q}=(0,0)$. This gap is closed at any temperature for any set of exchange parameters.

The structure factor has a peak at zero point only in the region of purely FM short-range order not discussed here. In all other cases, particularly for $(q,0)$ helix zero spin gap at trivial point does not generate corresponding $c_{\mathbf{q}}$ peak. In the  bottom row in Fig.~\ref{sopli} the side $c_{\mathbf{q}}$ peaks near the trivial point are generated by spectrum narrow dips at points $(q,0)$ being the traces of $(q,0)$ quantum helical order.

Let us mention that the circle-like vanishing spin gap, or in other words, the circle-like $c_{\mathbf{q}}$ maxima is the precursor of Brazovskii transition.~\cite{Brazov75_SJoEaTP}

Note in addition, that the correlation length of the FM circular state is much larger than the correlation length of the AFM circular state (compare Fig.~\ref{gyrate_AFM} and Fig.~\ref{sopli}).

\section{Conclusions}
To conclude, we have considered the topical case of the systems with multiple frustrating exchanges --- $S = 1/2$ two-dimensional $J_1-J_2-J_3$ Heisenberg model. Many important results for this problem are scattered over the localized islands of parameters, while areas in between still require investigation.

We use the spherically symmetric self-consistent approach for spin-spin Green's functions. It conserves all the key symmetries of the problem (the $SU(2)$-spin symmetry and the translational invariance) and strictly holds the characteristic limitation of low-dimensionality.

Note that purely analytical approaches (e.g. spin waves), in 2D should be used with caution (see~\cite{Schmid17_PR} for a recent review). Generally accepted results here are still absent, and as a rule, analytical approach serves as a basis for further numerical studies.
Let us underline that the problem at hand (detailed investigation of the large area of parameters) is resource consuming for direct numerical simulation: the frustration especially multi-exchange increases the well-known ``sign problem'' and there are also system size limitations. The method we use here allows to bypass these problems with the cost of some uncertainty related to the accuracy of multi-spin Greens-function approximation and to obtain in a reasonable time a physical picture in a very wide range of parameters for a moderate amount of processor-hours.

To be more specific, the evolution of a spin fluid between ferromagnetic and antiferromagnetic short-range order structures has been studied in the present work. We have found not only the structure factor but also the spin excitations spectra. It should be noted that recently circular structures have been detected by numerical methods for both ferromagnetic and antiferromagnetic signs of the first exchange~\cite{Seabra16_PRB,Reuthe11_PRB, Iqbal16_PRB}. We reproduce quantitatively most of the results from the mentioned papers. An isotropic FM helicoid was obtained in~\cite{Seabra16_PRB} for parameters $J_1 = -0.8$, $J_2 = 0.6$, $J_3 = 0.2$ (in our notations) and the SSSA approach reproduces this result. Similarly, SSSA reproduces Fig.~7 of~\cite{Reuthe11_PRB} and Fig.~1(b-e) of~\cite{Iqbal16_PRB}.

The structure factor and the spin excitations spectra (in particular, the circular structures) were investigated here on the line $J_3 = 0.17$ ($\psi = 10^{\circ}$). However, our calculations show that in the entire range $J_3 \in [0, 0.5]$ ($\psi \in [0^{\circ},  30^{\circ}]$), circular structures also appear.

We believe that our investigation in the framework of a physically transparent approach, combining a large number of limiting cases, complements and enriches this picture. The results obtained can be further refined in the most interesting areas by direct numerical methods.

It should be added that in the considered case of low but nonzero temperature, the spin state for any set of parameters is a singlet spin-liquid without long-range order. Our consideration refines the structures of a circular form --- quantum helical isotropical states.

The circular state is a continuous quantum superposition of helical states; the manifold of helices directions fills the circle-like curve.
The token of a circular state is a tube-like form of $c_{\mathbf{q}}$ and the circle-like manifold of spectrum $w_{\mathbf{q}}$ local minima. These key features enriched with traditional $c_{\mathbf{q}}$ and $w_{\mathbf{q}}$ parts lead to the zoo of peculiar spectra and structure factors.

Finally, we have investigated wide areas in the phase diagram when one local order state of spin-liquid transforms into another one. The nontrivial circular states are located just in the borderlands.

\section{Acknowledgments}

This work was supported by the Russian Foundation for Basic Research (project No. 19-02-00509). We express our gratitude to Russian Science Foundation (project No. 18-12-00438) for support of the numerical calculations. This work was carried out using supercomputers at Joint Supercomputer Center of the Russian Academy of Sciences (JSCC RAS), the Ural Branch of RAS, and the federal collective usage center Complex for Simulation and Data Processing for Mega-science Facilities at NRC “Kurchatov Institute”, http://ckp.nrcki.ru/.

\section*{References}
\bibliographystyle{iopart-num}
\bibliography{Biblio}
\end{document}